\documentclass[onecolumn]{aastex631}

\newcommand{\gsim}{\lower.5ex\hbox{$\; \buildrel > \over \sim \;$}}
\newcommand{\lsim}{\lower.5ex\hbox{$\; \buildrel < \over \sim \;$}}

\submitjournal{the ApJ}

\shorttitle{Shock wave}
\shortauthors{Zhao,Li,Tang et al}

\begin{document}


\title{Shock waves in the magnetic reconnection in the flares on the accretion disk of the SGR~A*}

\author{Tian-Le Zhao}
\affiliation{Center For Astrophysics, Guangzhou University, Guangzhou, Guangdong, 510006, People's Republic of China}
\affiliation{Great Bay Center, National Astronomical Data Center, Guangzhou, Guangdong, 510006, People's Republic of China}
\affiliation{Astronomy Science and Technology Research Laboratory of Department of Education of Guangdong Province, Guangzhou 510006, China}
\affiliation{CAS Key Laboratory for Research in Galaxies and Cosmology, Department of Astronomy, University of Science and Technology of China, Hefei 230026, China}
\email{tlzhao@gzhu.edu.cn}



\author[0000-0002-9093-6296]{Xiao-Feng Li}
\affiliation{School of Computer Science and Information Engineering, Changzhou Institute of Technology, Changzhou, Jiangsu, 213032, People's Republic of China}
\email{lixiaofeng@czu.cn}

\author{Ze-Yuan Tang}
\affiliation{Center For Astrophysics, Guangzhou University, Guangzhou, Guangdong, 510006, People's Republic of China}
\affiliation{Great Bay Center, National Astronomical Data Center, Guangzhou, Guangdong, 510006, People's Republic of China}
\affiliation{Astronomy Science and Technology Research Laboratory of Department of Education of Guangdong Province, Guangzhou 510006, China}
\affiliation{CAS Key Laboratory for Research in Galaxies and Cosmology, Department of Astronomy, University of Science and Technology of China, Hefei 230026, China}
\email{tangzy@mail.ustc.edu.cn}

\author{Kumar Rajiv}
\affiliation{Harish-Chandra Research Institute, Prayagraj 211019, Uttar Pradesh, India }
\email{rajivkumar@hri.res.in}
\correspondingauthor{Tianle Zhao, XiaoFeng Li, ZeYuan Tang, Kumar Rajiv}

\begin{abstract}

Sgr~A* often shows bright, episodic flares observationally, the mechanism of the flares intermittent brightening is not very clear. Many people believe the flares may formed by the non-thermal particles, which can be a consequence of the magnetic reconnection and shock waves. In this work, we use the larger magnetic loop in the presence of pseudo-Newtonian potential which mimics general relativistic effects. The simulation results show that the reconnection of magnetic field lines passes through a current sheet, which bifurcates into two pairs of slow shocks. We also find the shock waves heat the plasma, especially when the plasma density is low. The shock wave heating effect by the magnetic reconnection is confirmed by the simulation results, and thus the process of instantaneous brightening of the flares on the accretion disk can be explained.

\end{abstract}



\section{Introduction} \label{sec:introduction}

Sgr~A* is the super massive black hole resides at the center of our galaxy. It is a low-luminous and low accretion rate AGN, and thus it accretes the surrounding gas with a low accretion rate because of insufficient gas supplication \citep{2014ARA&A..52..529Y}. The Sgr~A* often shows bright, episodic $X$-ray and near-infrared flares observationally, and the most famous is {\it Very Large Telescope Interferometer} (VLTI) observed several near-infrared superflares which might be from the hot spots at about 10 to 15 Schwarzschild radii from Sgr~A* in 2018 \citep{2018A&A...618L..10G,2020ApJ...891L..36G}.  Some research showed the flares could be consistent with hotspots 
in the inner orbit of the AGN \citep{2021ApJ...917....8B, 2020MNRAS.497.4999D, 2021MNRAS.502.2023P}. The mechanism of the flares intermittent brightening is not very clear nowadays, and many people believe that the flares may form by the non-thermal particals\citep{2015MNRAS.454.3283Y, 2023MNRAS.520.1271L}.\newline

The mechanisms for forming non-thermal particles mainly include magnetic reconnection processes, shock waves, and convection, all of which can occur simultaneously. Shock waves are common in astrophysics environments, and there are various astrophysical phenomena associating the shock waves, such as various outburst phenomena, shocks in astrophysical jets and winds\citep{2019MNRAS.483.2538F}. On the theoretical front, many groups of works extensively study the shock wave around black holes\citep{2018MNRAS.480.3446D}.
In addition, many researchers consider the magnetohydrodynamic (MHD) environment to study the existence of shock wave in accretion flow, which is also examined numerically\citep{2014MNRAS.441.2354O,2015MNRAS.453..147O,2017MNRAS.472.4327S,2005ApJ...625...60N,2006ApJ...645.1408T,2007ApJ...657..415F,2016ApJ...827...31F}. Some research also find the shock existence in the low angular momentum disc-like structure in hydrostatic equilibrium both in pseudo-Newtonian potential and in full relativistic approach\citep{2002ApJ...577..880D,2012NewA...17..254D,2015MNRAS.447.1565S,2021ApJ...910....9K}. \citet{2017MNRAS.469.4221K} analyze flows around a rotating black hole and the result has shown the shocks in both accretion flow and outflows (jets/winds). \citet{2007ApJ...661..416L,2017MNRAS.465.1409L,2020MNRAS.491.4194L} also do the simulation work on the particle acceleration across the shocks. \citet{2020ApJ...897...64Y} use the MHD method to study the eruptive solar flares and found that shock waves can heat plasmas above the loop-top to a higher temperature.\newline

This simulation study involves investigating the shock wave that heats the plasma during magnetic reconnection within the flares of SgrA*. Many studies focus on the global disk of the Sgr~A* with the GRMHD code\citep{2014MNRAS.441.2354O,2017MNRAS.472.4327S}. In this work, we do the high-resolution local simulation to study the shock wave during the magnetic reconnection. We select the small simulation region on the surface of the hot accretion disk. Section 2 gives a very brief introduction to the numerical model. The ensuing section, \S 3, shows the case of numerical results, while the main conclusions are summarized in Section \S 4.\newline

\section{Numerical model} \label{sec:model}
This work can be described by the  single-fluid MHD equations including the pseudo-Newtonian potential, magnetic diffusion and thermal conduction, and the equations are written as follows\citep{2017ApJ...841...27N,2018RAA....18...45Z,2020MNRAS.499.1561Z,2021MNRAS.508.5251Z,2023MNRAS.526.1198Z}:
\begin{eqnarray}
	\partial_t \rho &=& -\nabla \cdot \left(\rho \mathbf{v}\right),                                                                \\ 
	\partial_t \mathbf{B} &=& \nabla \times \left(\mathbf{v} \times \mathbf{B}-\eta\nabla \times \mathbf{B}\right), \label{e:induction}\\
	\partial_t (\rho \mathbf{v}) &=& -\nabla \cdot \left[\rho \mathbf{v}\mathbf{v}
	+\left(p+\frac {1}{2\mu_0} \vert \mathbf{B} \vert^2\right)\mbox{\bfseries\sffamily I} \right] +\nabla \cdot \left(\frac{1}{\mu_0} \mathbf{B} \mathbf{B} \right) + \rho \nabla \phi, \\
	\partial_t e &=& - \nabla \cdot \left[ \left(e+p+\frac {1}{2\mu_0 }\vert \mathbf{B} \vert^2\right)\mathbf{v} \right] +\nabla \cdot \left[\frac {1}{\mu_0} \left(\mathbf{v} \cdot \mathbf{B}\right)\mathbf{B}\right] + \nabla \cdot \left[ \frac{\eta_{TD}}{\mu_0} \textbf {B} \times \left(\nabla \times \mathbf{B}\right) \right] \nonumber\\
	& & -\nabla \cdot \mathbf{F}_\mathrm{C}+\rho \mathbf{g} \cdot \mathbf{v},   \\
	e &=& \frac{p}{\gamma-1}+\frac{1}{2}\rho \vert  \mathbf{v} \vert^2+\frac{1}{2\mu_0}\vert \mathbf{B} \vert^2,   \\
	p &=& \frac{2\rho k_B T}{m_i}, \\
	\phi&=&-GM/(\sqrt{x^2+y^2}-R_s),       
\end{eqnarray}
where $\mathbf{B}$, $\rho$, $v$, $p$, $e$, $\eta_{TD}$, $F_C$ represent the magnetic field, plasma density, fluid velocity, gas pressure, energy density, magnetic diffusion, and thermal conduction respectively. This time, the gravity is considered in both $x$-direction and $y$-direction. The constants $\mu_0$, $k_B$. $m_i$  $M$, $G$, $R_s$ and $\gamma=5/3$ are the permeability of vacuum, Boltzmann constant, the mass of proton, the mass of the black hole, the gravitational constant, the Schwarzschild radius, and the adiabatic index, respectively \citep{2017ApJ...841...27N,2018RAA....18...45Z,2020MNRAS.499.1561Z,2021MNRAS.508.5251Z,2023MNRAS.526.1198Z}.  

\begin{figure*}
	\centering
	\includegraphics[width=0.4\textwidth]{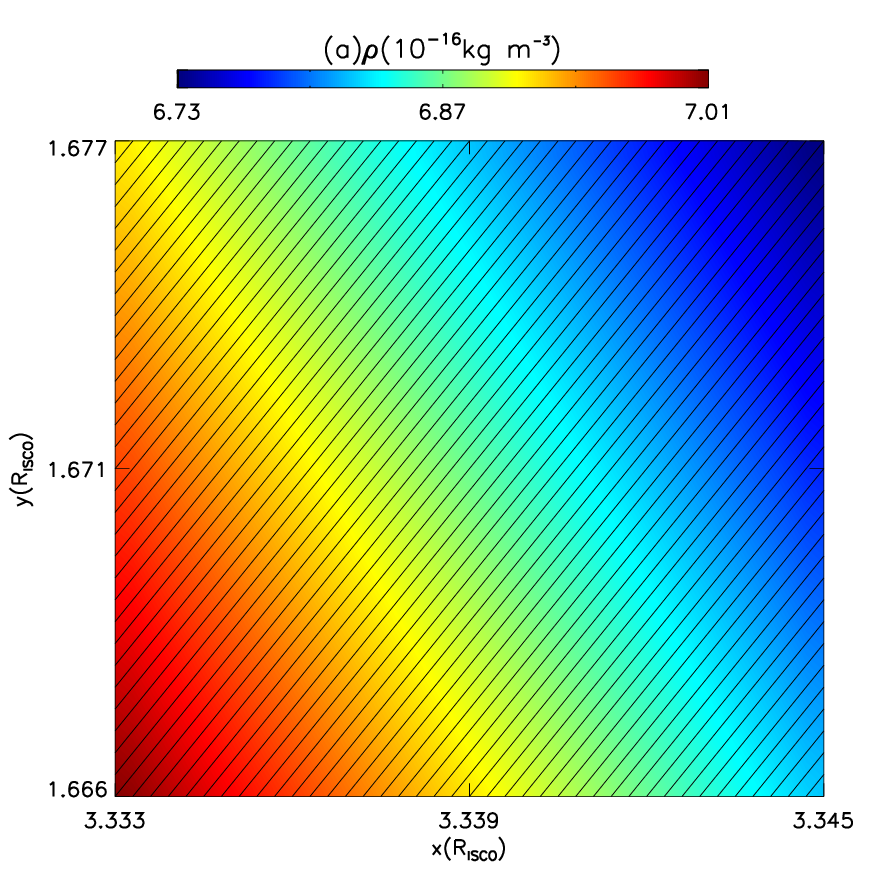}
	\includegraphics[width=0.4\textwidth]{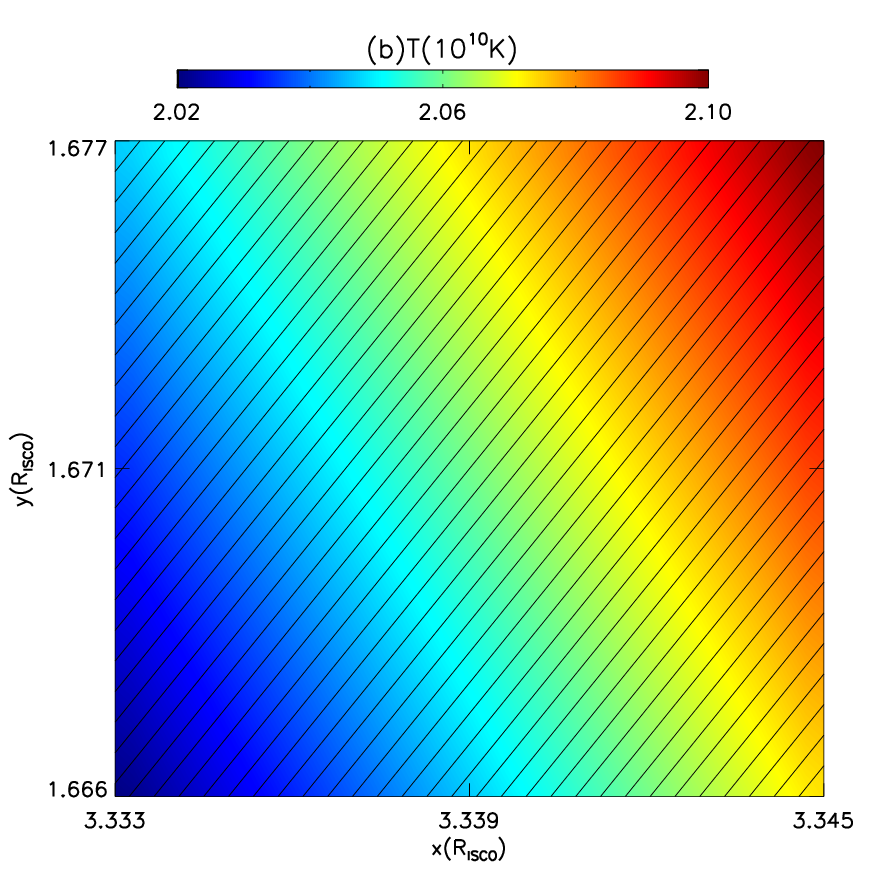}
	\caption {The initial distribution of density and temperature and configuration of the magnetic field distribution at $t=0$~s. The solid black lines represent the magnetic fields.
		\label{den}}
\end{figure*}

The pseudo-Newtonian potential we use can accurately describe most of the relativistic properties of space-time \citep{1980A&A....88...23P}, except some of the properties which are effective very close to the black hole, such as the ray deflection caused by general relativity, however, it is not strong in the chosen region of this study. 
In this work, the 2-D Cartesian coordinates (x, y) are used following \citep{2020MNRAS.499.1561Z,2023MNRAS.526.1198Z}, and the simulation area is [$R_{in}$,$R_{in}+400L_0$]$\times$[$H_0$,$H_0+400L_0$], with $R_{in}=10R_s$, $H_0=5R_s$, $L_0=10^6$~m, which is larger than in \citet{2023MNRAS.526.1198Z}. In this paper, the mean free path in low mass accretion rate flow is quite high \citep{2019AIPA....9j5307L}. However, we use a much higher anomalous magnetic diffusivity to simulate the magnetic reconnection\citep{2020ApJ...888..104L}, and thus the mean free path used in this simulation is in millimeter scale. The simulation size is much larger than the mean free path used in this simulation, Larmor radius, and Debye length. We considered the accretion disk to be hot and geometrically thick. The simulation region is located in the environment of the magnetically threaded accretion disk that has inclined magnetic field lines above the disk surface which can be closely related to the assumptions of these studies \citep{2009MNRAS.395.2183Y,2011ApJ...737...94C,2012MNRAS.426.3241N}. The initial conditions are set as follows. We assume that the gas density varies along the radial direction that increases towards the center which follows qualitatively the variation of density in accretion flows as also used in these studies \citep{2018RAA....18...45Z,2020MNRAS.499.1561Z,2023MNRAS.526.1198Z}. The gas density distribution is $\rho(x, y)=\rho_0\exp(-(x^2+y^2)/H_0^2)$, and $\rho(R_{in},H_0)=1.04\times10^{-13}$~$\mathrm{kg}$~$\mathrm{m^{-3}}$ is taken, which is shown in Fig \ref{den}(a). The gas pressure is determined by the hydrostatic equilibrium formula $\rho\nabla \phi=-\nabla p$ with $p(R_{in}, H_0)=2\rho(R_{in}, H_0)k_BT_0/m_i$ and $T_0=2\times10^{10}$~K, which can also gives initial temperature distribution   as shown in Fig \ref{den}(b). 
The Spitzer anisotropic heat conduction flux $F_C$ (e.g., see also Spitzer 1962)  can lead to the transfer of the heat from the reconnection X-point into the O-point in the plasmoids, and this process can increase the rate of reconnection and the efficiency of the conversion of magnetic energy into the thermal energy and kinetic energy of the bulk motions\citep{2012ApJ...758...20N} and the temperature-dependent magnetic diffusivity $ \eta_{TD} $ are set as:
\begin{equation}
	\eta_{\mathrm{TD}}=10^8\left(T_0/T\right)^{3/2}+10^9\left[1-\mathrm{tanh}\left(\frac{y-2L_0}{0.2L_0}\right)\right](\mathrm{m^2 s^{-1}}),
	\label{eta}
\end{equation}

which is the same as those in our previous work\citep{2017ApJ...841...27N,2018RAA....18...45Z,2020MNRAS.499.1561Z,2021MNRAS.508.5251Z,2023MNRAS.526.1198Z}, which can enhance the magnetic reconnection in the simulation. The initial background magnetic field is set as $B_{x0}=-0.6b_0$ and $B_{y0}=-0.8b_0$, with $b_0=32$~G. The magnetic field strength of 32 G near the black hole is consistent with theoretical and simulations-based understandings  \citep[$16-100$~G,][]{2009MNRAS.395.2183Y,2020MNRAS.497.4999D,2021ApJ...917....8B}, and also satisfies constraints from Event Horizon Telescope observations \citep[average 28 G at $3.65r_{\rm s}$][]{}\citep{2024ApJ...964L..26E}. The outflow condition is used in the left, right, and upper boundaries, and the magnetic fields in the bottom layer with ghost grid cells are set as follows\citep{2017ApJ...841...27N}:
\begin{equation}
	b_{xb}=-0.6b_0+\frac{100L_0(y-y_0)b_1 f}{[(x-x_0)^2+(y-y_0)^2]}\left\lbrace 
	\left[\tanh\left(\frac{x-170L_0-R_{in}}{\lambda}\right)-\tanh\left(\frac{x-330L_0-R_{in}}{\lambda}\right)\right] \right\rbrace   , 
\end{equation}
\begin{equation}
	b_{yb}=-0.8b_0-\frac{100L_0(x-x_0)b_1 f}{[(x-x_0)^2+(y-y_0)^2]}\left\lbrace
	\left[\tanh\left(\frac{x-170L_0-R_{in}}{\lambda}\right)-\tanh\left(\frac{x-330L_0-R_{in}}{\lambda}\right)\right]\right\rbrace    , 
\end{equation}
with $x_0=R_\mathrm{in}+250L_0$ and $y_0=H_0-10L_0$, $\lambda=0.5L_0$, $b_1=6.4$~G and $ t_1=300 $~s. The factor $f$ is $f=t/t_{1}$ for $t\leqslant t_{1}$ and 1 for $t>t_1$. \citet{2009MNRAS.395.2183Y} shows the emerging magnetic field (close field) and open field on the geometrically accretion disk of the Sgr A*, magnetic reconnection occurs between them. In this study, we set the emerging flux loop from the accretion disk is larger than the previous study \citep{2023MNRAS.526.1198Z}. The adaptive mesh refinement method in this simulation is the same as in the paper by \cite{2017ApJ...841...27N}, and the grid used in this paper is $(40\times2^9) \times (40\times2^9)$.\newline

\section{Numerical results} \label{sec:result}
\begin{figure*}
	\centering
	\includegraphics[width=0.75\textwidth]{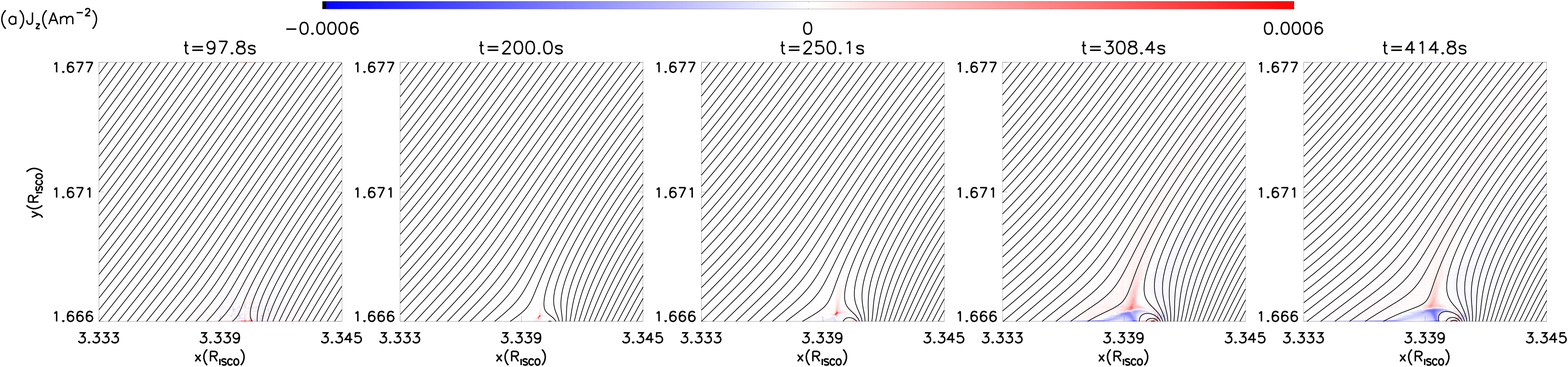}
	\includegraphics[width=0.75\textwidth]{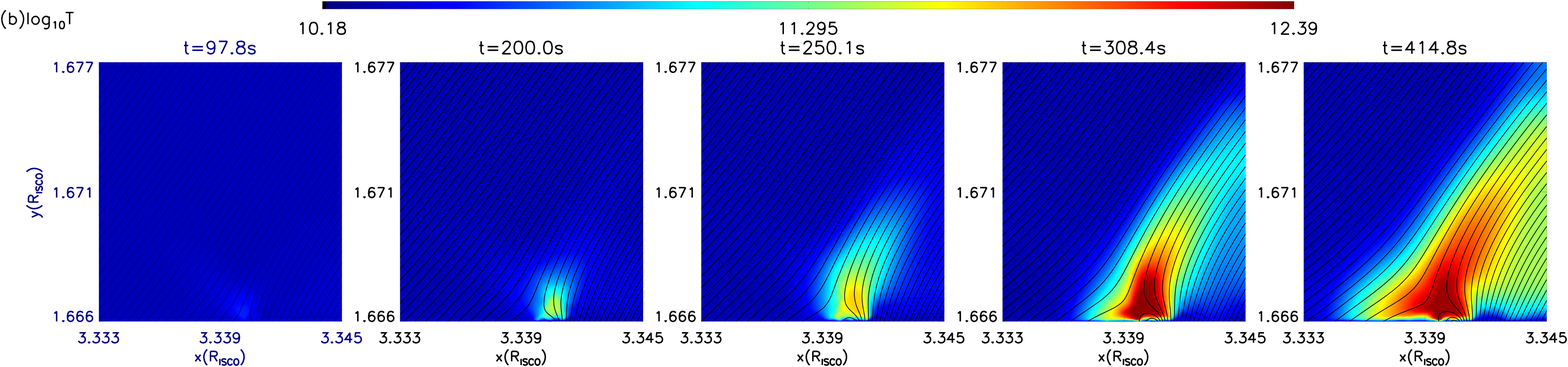}
	\includegraphics[width=0.75\textwidth]{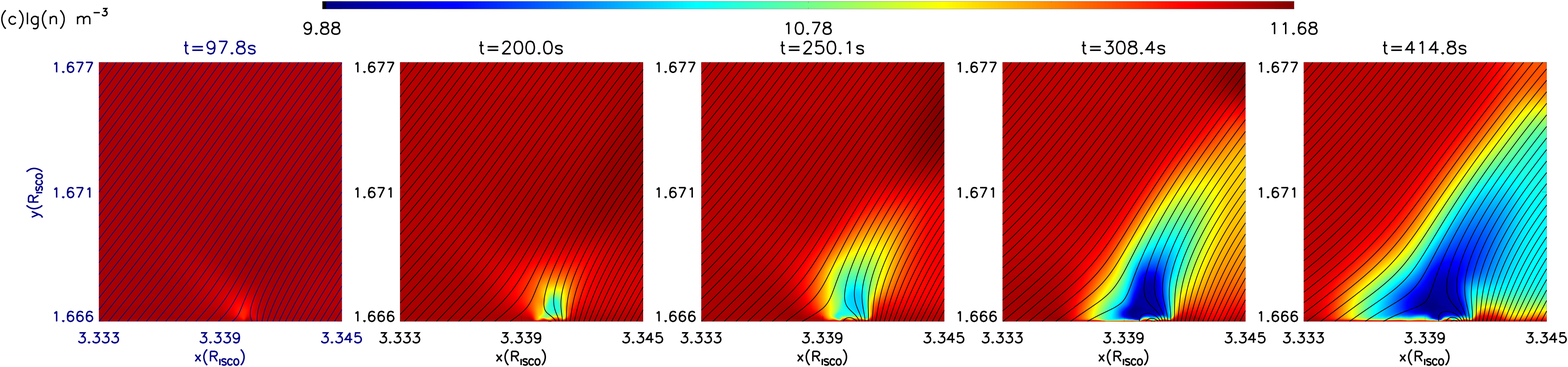}
	\includegraphics[width=0.75\textwidth]{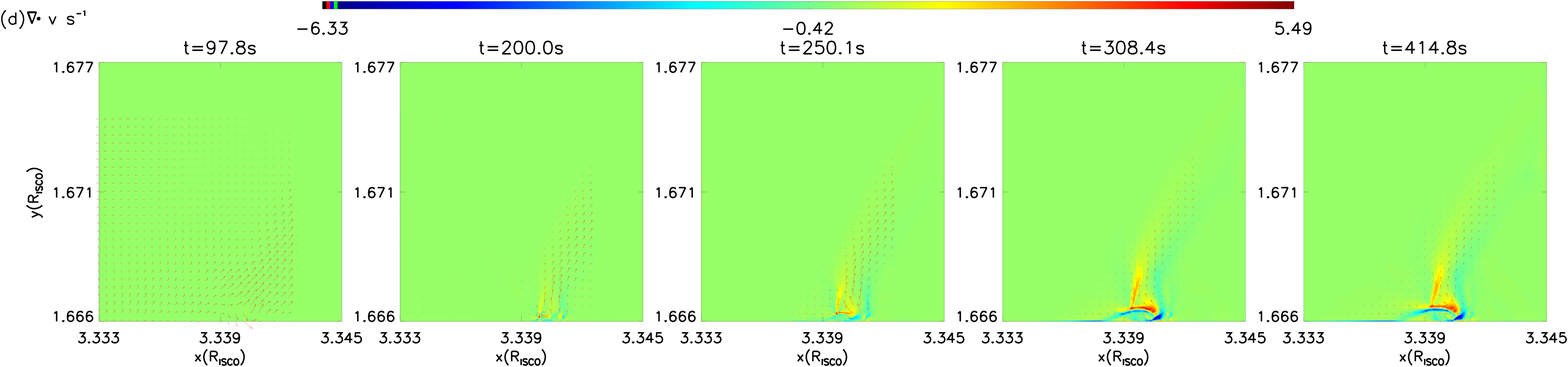}
	\caption {Distributions of the different variables at five different times in our simulation. (a) Current density, $J_z$, the red arrows show the direction of the magnetic field line, (b) Log of Temperature, $ \mathrm{log}_{10}(T) $, (c) Log of particle number density, $\mathrm{log}_{10}(n) $, (d) Velocity divergence, $\nabla\cdotp v$ with the velocity vector.
		\label{all}}
\end{figure*}

Figure \ref{all}(a)-(d) represent the current density, temperature, density, and velocity divergence, respectively, and each column in Figure 2 corresponds to the time $t=97.8$~s, $t=200.0$~s, $t=250.1$~s, $t=281.9$~s and $t=308.4$~s from the beginning. 
The values of flow variables are represented by the color bar at the top of each row.
Figure \ref{all}(a) shows that the current sheet rises in the simulation area. Figure  \ref{all}(b) and  \ref{all}(c) shows that the plasma is gradually heated near the outflow region, especially with low plasma density. As is shown in Figure \ref{all}(d), the magnetic lines reconnect when passing the current sheet, and bifurcate into two pairs of shocks. As is described in \citet{2014masu.book.....P}, at the end of the current sheet, there is a pair of slow-mode shock waves that propagate and remain as standing waves in a steady flow, which is similar to the Petschek-type magnetic reconnection. So the shocks shown in Figure \ref{all}(d) are slow mode shocks. Figure \ref{all}(d) also shows that the plasma on the right is heated by the shock waves\citep{2020ApJ...897...64Y,2022RAA....22h5010Z}. The comparison of the last two columns shows that the system changes only a little after the newly emerging magnetic field stops.\newline

\begin{figure*}
	\centering
	\includegraphics[width=0.75\textwidth]{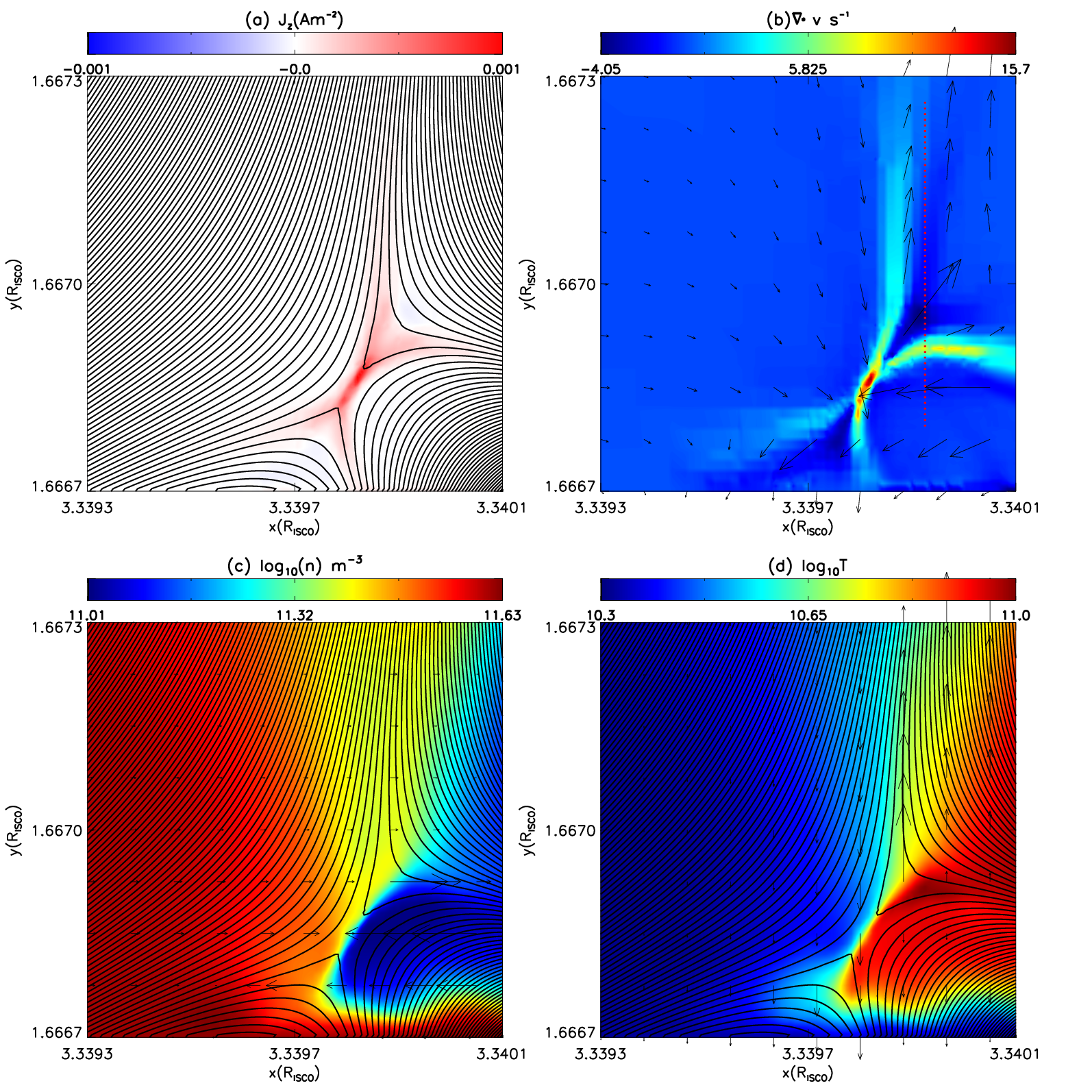}
	\caption{Physical quantity distribution for the shock wave during the magnetic reconnection at $t=194.39$~s. (a)current density $J_z$ shows current sheet in red color, (b)velocity divergence $\nabla\cdotp v$, (c)particle number density $\mathrm{log}_{10}(n) $, (d)temperature $\mathrm{log}_{10}(T)$.
		\label{shockwave}}
\end{figure*}

\begin{figure*}
	\centering
	\includegraphics[width=0.35\textwidth]{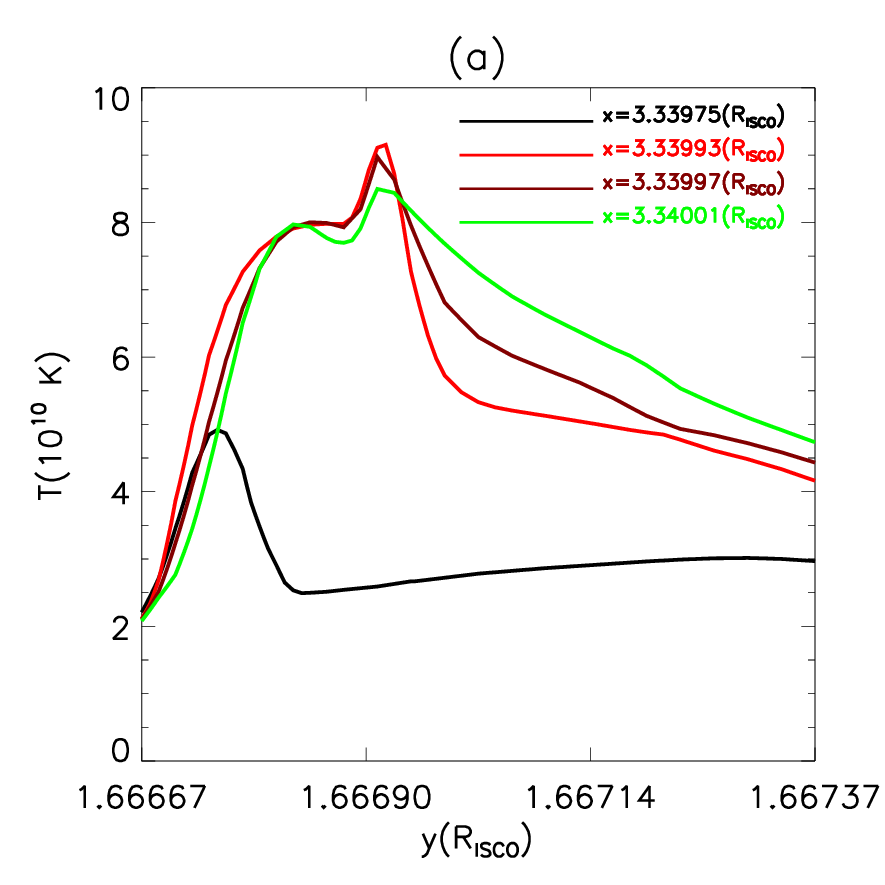}
	\includegraphics[width=0.35\textwidth]{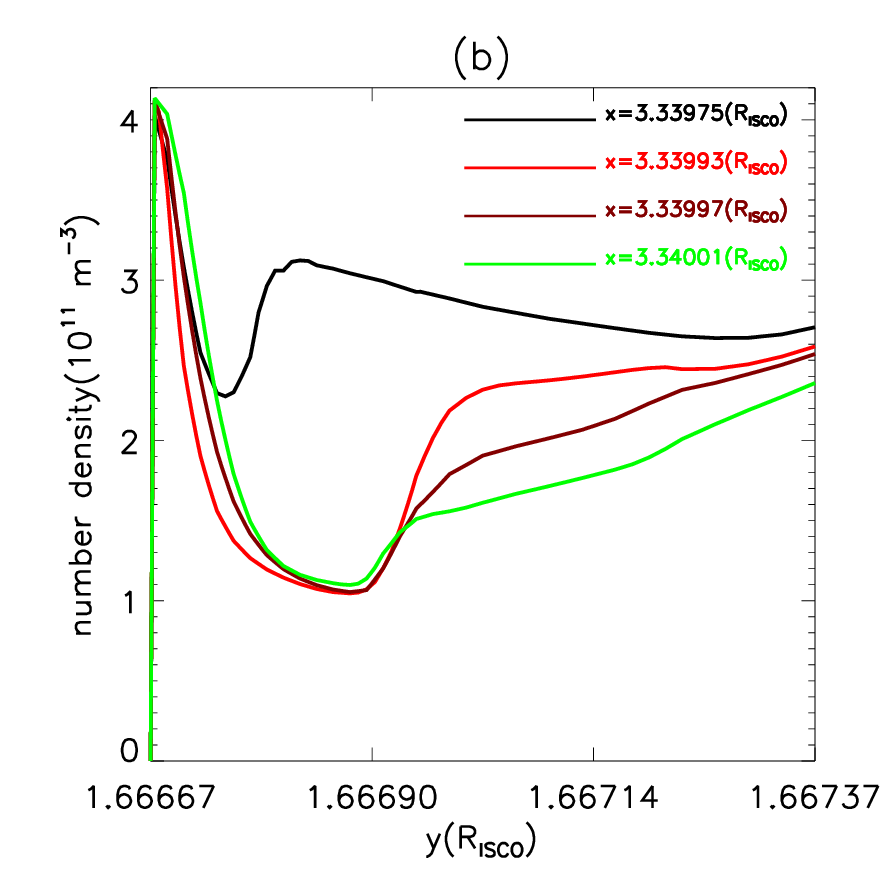}
	\includegraphics[width=0.35\textwidth]{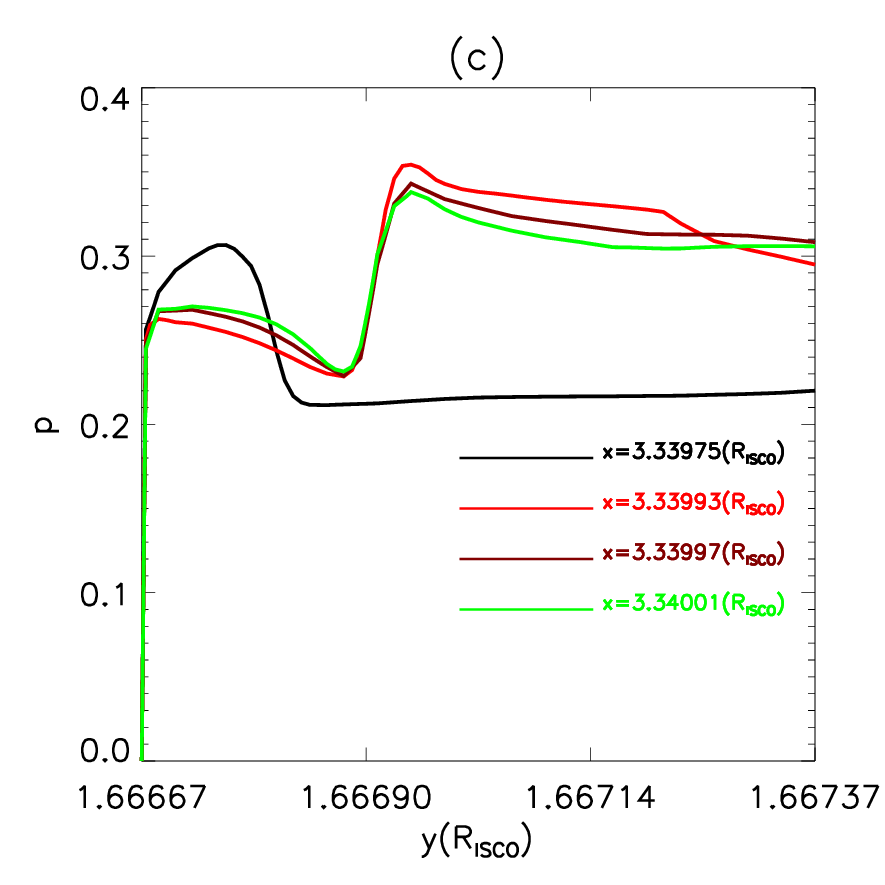}
	\includegraphics[width=0.35\textwidth]{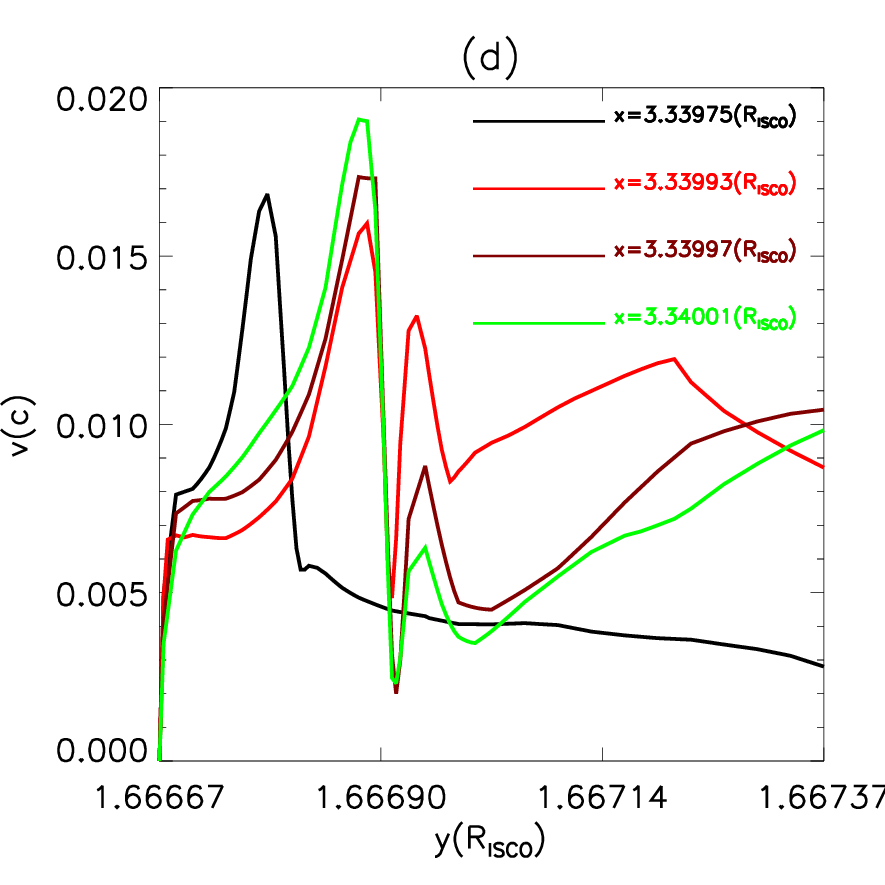}
	\caption{Physical quantity distribution vary along $y$-direction at four different $x$ location at $t=194.39$~s.(a) Plasma temperature $T$ and $v_n$, (b)Number density of particles $n$, (c) gas pressure $p_{gas}$, (d) plasma speed $v$.
		\label{tv4}}
\end{figure*}

\begin{figure*}
	\centering
	\includegraphics[width=0.35\textwidth]{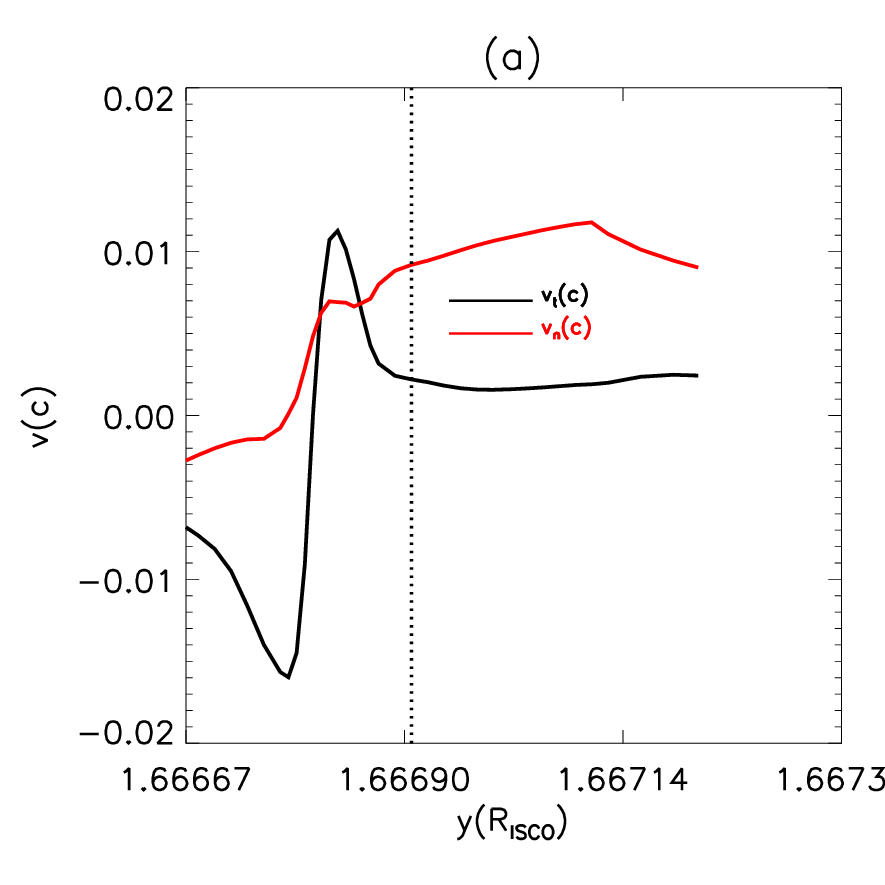}
	\includegraphics[width=0.36\textwidth]{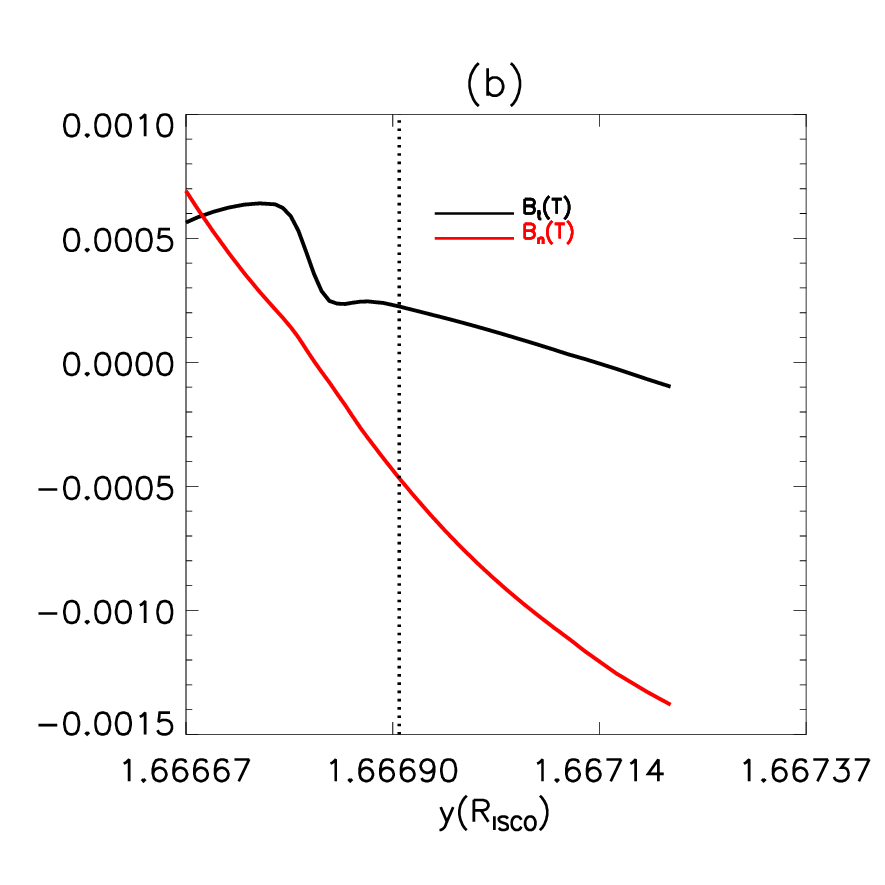}
	\includegraphics[width=0.36\textwidth]{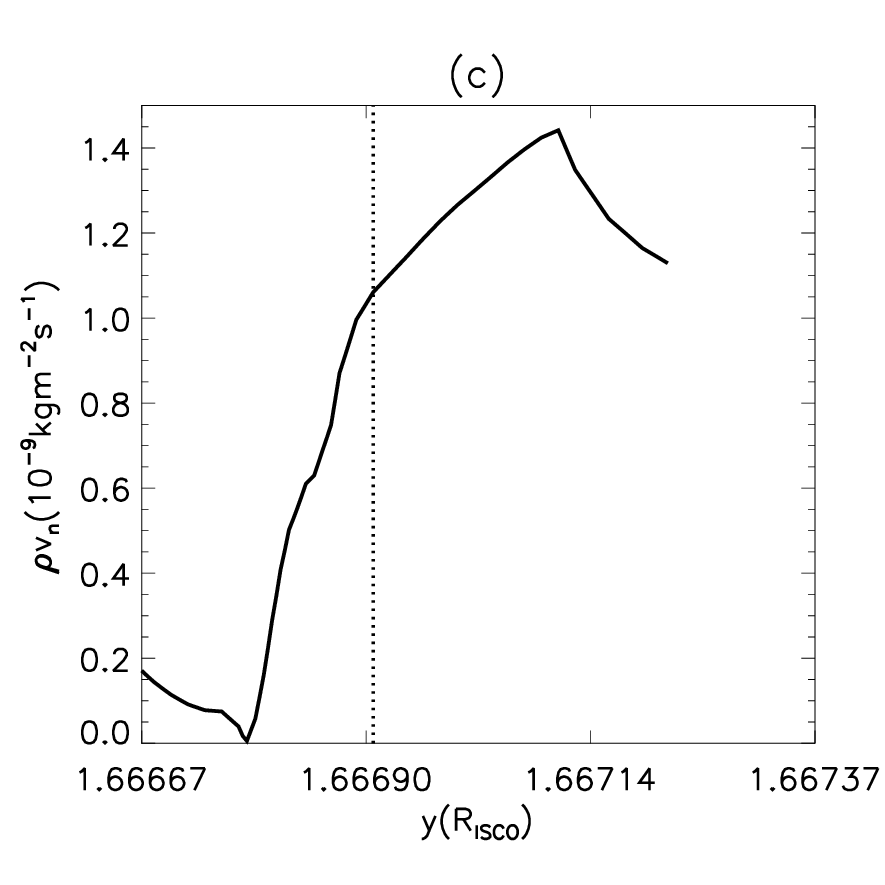}
	\includegraphics[width=0.35\textwidth]{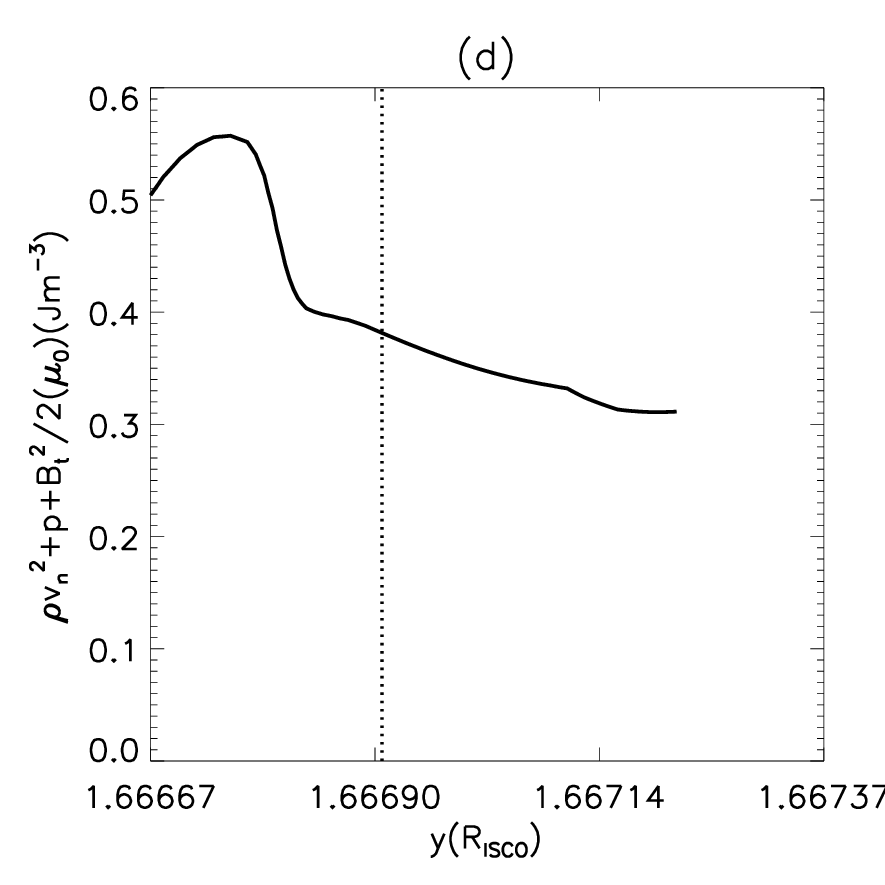}
	\caption{Physical quantity distribution vary along $y$-direction along the dot line at Fig \ref{shockwave} position at $t=194.39$~s.(a) Two components of the plasma velocities, $v_t$ and $v_n$, (b) Two components of the magnetic field $B_t$ and $B_n$, (c) variable $\rho v_n$, (d)variable $\rho v^{2} +p +\frac{B^{2}}{2\mu_{0}}$.
		\label{tv}}
\end{figure*}

To investigate the structure of the shock wave more clearly,  we run the highest AMR level case $(160\times2^{15}) \times (160\times2^{15})$. Unfortunately, the highest level case requires a large amount of calculation resources, so we only run this case until the appearance of the shock waves.The shock wave structure and the physical quantities are shown in Fig \ref{shockwave}-\ref{tv}.
Figure \ref{shockwave} (a)-(d) shows the distributions of current density, velocity divergence, particle density, and temperature respectively at the representative moment t=194.39s, when the shock waves appeared, and the vectors in Figure \ref{shockwave} (b)-(d) represent the velocity vector, and in $x$ and $y$ direction, respectively. Figure \ref{shockwave} (a) presents a structure of the positive current sheet pointing to the upper right direction, and Figure \ref{shockwave} (b) shows there exists a shock wave structure at both ends of the current sheet, especially in the upper region. Figure \ref{shockwave} (c) and (d) present a shock wave heating region, especially in the outflow region where a large quantity of low density plasma is heated greatly.\newline

Figure \ref{tv4} shows the jump along the $y$-direction at different $x$-location. The black solid line $x=3.33975R_\mathrm{ISCO}$ passes through the shock region at the left side of the current sheet (the small red region in panel b of Fig. \ref{shockwave}) in the simulation area, while the other three solid lines $x=3.33993R_\mathrm{ISCO}$(the red solid line), $x=3.33997R_\mathrm{ISCO}$(the brown solid line) and $x=3.34001R_\mathrm{ISCO}$ (the green solid line) pass through the shock wave at the right side of the current sheet. The temperature jumps can be seen at the top part of the three curves in Figure \ref{tv4} (a), which indicates that the plasma is greatly heated by the shock wave. 
Figure \ref{tv}(b) shows the number density of the particles decreases when passing through the shock wave and increases in the post-shock region. 
Figure \ref{tv4}(c) shows the gas pressure jumps at the shock location, which indicates that the plasma is compressed by the shock wave.
Figure \ref{tv4}(d) shows the plasma speed variations along the y-direction. Here the plasma speeds have two peaks in three curves (red, brown, and green). At the first peak, the speed increases and then sharply decreases at the location of the shock wave, and corresponding jumps in other quantities are shown in the other plots of this figure. At the second peak, the plasma speeds of those three curves (red, brown, and green) have similar rising and decreasing slopes. Corresponding to these variations in the speed, jumps are not found in the other plots of gas pressure, temperature, and density. Therefore, the speed variations in the second peak cannot represent the shock wave.
The jumps of several physical quantities in Figure \ref{tv4} allow us to see the structure of the shock wave much more clearly. These shock wave phenomena can also provide energy for intermittent brightening of flares due to sudden heating.\newline

We analyze the variables that change along the direction perpendicular to the shock wave surface. The MHD conditions are $B_{n1}=B_{n2}$, $\rho_{1} v_{n1}=\rho_{2} v_{n2}$, and $\rho_{1} v_{n1}^{2} +p_{1} +\frac{B_{t1}^{2}}{2\mu_{0}}=\rho_{2} v_{n2}^{2} +p_{2} +\frac{B_{t2}^{2}}{2\mu_{0}}$, the subscript $t$ represents the tangential component to the shock front and the subscript $n$ represents the normal component to the shock front\citep{2014masu.book.....P,2017ApJ...841...27N}. Figure \ref{tv} shows the jumps along the dotted line via $y$-direction. Figure \ref{tv}(a) shows the two components of the plasma velocity $v_t$ and $v_n$, which shows the value varies significantly when passing through the shock wave. Figure \ref{tv}(b) shows the two components of the magnetic field $B_t$ and $B_n$, both of which decrease before and after the shock wave. However, the tangential magnetic component decreases sharply at the shock location, which is one of the features of the slow mode shock.
Figure \ref{tv}(c)(d) show the distribution of the $\rho v_n$ and $\rho v_{n}^{2} +p +\frac{B_{t}^{2}}{2\mu_{0}}$ along the dotted line before and after passing through the shock wave, respectively. The changes of the $B_{n}$, $\rho v_{n}$ and $\rho v_{n}^{2} +p +\frac{B_{t}^{2}}{2\mu_{0}}$ are not very significant so that they can be ignored. One should point out that the theoretical analysis is idealized and it assumes all the physical variables in the MHD jump conditions are time-independent and uniform at each side of the shock front\citep{2017ApJ...841...27N}. However, in the numerical simulations, all the variables change with time and location. Therefore, there should be some deviations between the analytical and numerical results. \newline

\section{Discussion and Conclusion}
In this work, based on the magnetic reconnection in the \citep{2020MNRAS.499.1561Z,2021MNRAS.508.5251Z,2023MNRAS.526.1198Z}, we set a larger magnetic emerging loop and magnetic resistivity with a larger simulation box compared to our previous works.  This time, instead of assuming constant gravity, we also considered the variation of gravity in both the $x$ and $y$-directions. 
So we believe that the combined effects due to the addition/enhancement of those things can facilitate the conditions for the shock formation in the magnetic reconnection region on the surface of the fluid collision. This can be due to the generation of extra forces in the medium, so the shock features that have not been seen in our previous studies become prominent in this study.
The main focus of this numerical simulation is the formation of current sheets and shock structures through magnetic reconnection, as well as the heating of plasma by shock waves. Shock waves mainly heat the plasma in the outflow region, and the lower the density of the plasma, the more obvious the heating. In addition, when passing through a shock wave, the temperature increases sharply, while the velocity decreases instantly and then increases again. The numerical results confirm that the shock waves appear on the accretion disk by the magnetic reconnection and heat the plasma, which can explain the process of instantaneous brightening of the flares on the accretion disk.\newline

Different from the case in \citet{2023MNRAS.526.1198Z} using the constant gravity in $x$-direction, this work use the pseudo-Newtonian potential, and thus the current sheet can keep positive and point to the right. Because of the downward force, it is easy for the outflow region and the front fluid to collide and produce shock waves. Different from the model by \citet{2023A&A...672A..62A}, the jet base temeprature is $3.0 \times 10^{10}$~K and the density is $3.5 \times 10^6$~$\mathrm{m^{-3}}$. In this work, the initial particle number density in simulation area is $4.05 \times 10^5$-$4.2 \times 10^5$~$\mathrm{m^{-3}}$, the initial temperature is $2.01\times 10^{10}$-$2.10\times 10^{10}$~K, and the initial magnetic field is $32$~G, similar to the observing work in \citet{2022ApJ...931....7B}. In this work, the shock waves appear in the simulation area, but there is no blob structure during the manetic reconection. Next time we will set a more complicated model to explain the instant brightening mechanism of flares more comprehensively.\newline

\section*{acknowledgments}
This work is supported by China Postdoctoral Science Foundation (No.2023M730768) and National Natural Science Foundation of China (No. 12203014). The work is also partially supported by the Natural Science Foundation of Guangdong Province (2019B030302001). We acknowledge the science research grants from the China Manned Space Project, with NO. CMS-CSST-2021-A06. We also acknowledge support from Scientific and Technological Cooperation Projects (2020-2023) between the People’s Republic of China and the Republic of Bulgaria. We would like to thank the referee for the critical comments and suggestions, Prof. Yanrong Li, Prof. Feng Yuan, Prof. Feng Wang, Prof. Lei Ni, Prof. YeFei Yuan, Dr JiaWen Li, research assistant ShuKang Chen, ShaoWei Zhong for their helpful comments. The numerical calculations in this paper have been done on supercomputing system in the National Supercomputer Center in Guangzhou.

%





\bibliography{shockwave.bib}{shockwave.bib}
\bibliographystyle{aasjournal}

\end{document}